\documentclass{llncs}

\usepackage{epsfig}
\usepackage{citesort}
\usepackage{fancybox,epic}
\usepackage{latexsym,amssymb}

\newcommand{\alub}{\mbox{\sf Alub}}

\newcommand{\adom}{{\it ADom}}
\newcommand{\aint}{{\it AInt}}
\newcommand{\aatom}{{\it AAtom}}
\newcommand{\acert}{{\it ACert}}
\newcommand{\dom}{D_\alpha}
\newcommand{\inten}{I_\alpha}
\newcommand{\atom}{S_\alpha}
\newcommand{\cert}{Cert_\alpha}

\newcommand{\certdel}{Cert_\alpha^{del}}

\newcommand{\certadd}{Cert_\alpha^{add}}

\newcommand{\lsem}{\mbox{$\lbrack\hspace{-0.3ex}\lbrack$}}
\newcommand{\rsem}{\mbox{$\rbrack\hspace{-0.3ex}\rbrack$}}
\newcommand{\sqbrack}[1]{\lsem #1 \rsem}

\newcommand{\p}{{\sqbrack{P}}}

\newcommand{\semabs}{\p_\alpha}

\newcommand{\CP}{\mbox{\it CP}}
\newcommand{\defff}{\mathit{Def}}

\newcommand{\short}[1]{}

\long\def\comment#1{}

\newcommand{\ciaopp}{\texttt{CiaoPP}}

\newcommand{\entry}[2]{#1 \mapsto #2}

\newcommand{\dependency}[2]{#1 \Rightarrow #2}

\newcommand{\Mod}[1]{\mathit{{\it Upd}({#1})}}

\newcommand{\addition}[1]{\mathit{Add(#1)}}
\newcommand{\deletion}[1]{\mathit{Del(#1)}}

\newcommand{\analyzerA}{{\sc Analyze}}

\newcommand{\certificateA}{{\sf Cert}}

\newcommand{\eprog}{\it U_P} 

\newcommand{\certificateB}{{\sf In\-c\_\-Ce\-rt}}

\newcommand{\certificateL}{{\sf Ext\-\_\-Ce\-rt}}

\newcommand{\error}{{\sf Error}}

\renewcommand{\odot}{\oplus}

\begin{document}

\title{Some Issues on Incremental Abstraction-Carrying Code}

\author{Elvira Albert\inst{1} \and Puri Arenas\inst{1} \and Germ\'{a}n
  Puebla\inst{2}}

\institute{Complutense University of Madrid,
\email{\{elvira,puri\}@sip.ucm.es}
\and
Technical University of Madrid,
\email{german@fi.upm.es}
}

\maketitle

\begin{abstract}

  \emph{Abstraction-Carrying Code} (ACC) has recently been proposed as
  a framework for proof-carrying code (PCC) in which the code supplier
  provides a program together with an \emph{abstraction} (or abstract
  model of the program) whose validity entails compliance with a
  predefined safety policy. The abstraction thus plays the role of
  safety certificate and its generation (and validation) is carried
  out automatically by a fixed-point analyzer.  Existing
  approaches for PCC are developed under the assumption that the
  consumer reads and validates the entire program w.r.t.\ the
  \emph{full} certificate at once, in a non incremental way.  In this
  abstract, we overview the main issues on \emph{incremental} ACC. In
  particular, in the context of logic programming, we discuss both the
  generation of incremental certificates and the design of an
  incremental checking algorithm for untrusted \emph{update}s of a
  (trusted) program, i.e., when a producer provides a modified version
  of a previously validated program. By update, we refer to any
  arbitrary change on a program, i.e., the extension of the program
  with new predicates, the deletion of existing predicates and the
  replacement of existing predicates by new versions for them. We also
  discuss how each kind of update affects the incremental extension in
  terms of accuracy and correctness.

\end{abstract}

\section{Introduction}

Proof-Carrying Code (PCC) \cite{Nec97} is a general technique for
mobile code safety which proposes to associate safety information in
the form of a \emph{certificate} to programs. The certificate (or
proof) is created at compile time by the \emph{certifier} on the code
supplier side, and it is packaged along with the code.  The consumer
who receives or downloads the (untrusted) code+certificate package can
then run a \emph{checker} which by an efficient inspection of the code
and the certificate can verify the validity of the certificate and
thus compliance with the safety policy.  The key benefit of this
``certificate-based'' approach to mobile code safety is that the
consumer's task is reduced from the level of proving to the level of
checking, a task which should be much simpler, efficient, and automatic
than generating the original certificate.

Abstraction-carrying code (ACC) \cite{lpar04-ai-safety} has been
recently proposed as an enabling technology for PCC in which an
\emph{abstraction} (i.e., an abstract model of the program) plays the
role of certificate. An important feature of ACC is that not only the
checking, but also the generation of the abstraction (or fixpoint) is
\emph{automatically} carried out by a fixed-point analyzer.
Lightweight bytecode verification \cite{RR98} is another PCC method
which relies on analysis techniques (namely on type analysis in the
style of those used for Java bytecode verification \cite{JVM03}) to
generate and check certificates in the context of the Java Card
language. In this paper, we will consider analyzers which construct a
program \emph{analysis graph} which is interpreted as an abstraction
of the (possibly infinite) set of states explored by the concrete
execution.  Essentially, the certification/analysis carried out by the
supplier is an iterative process which repeatedly traverses the
analysis graph until a fixpoint is reached.  A key idea in ACC is
that, since the certificate is a fixpoint, a single pass over the
analysis graph is sufficient to validate the certificate in the
consumer side. 

Existing models for PCC (ACC among them) are based on checkers which
receive a ``certificate+program'' package and read and validate the
entire program w.r.t.\ its certificate at once, in a non incremental
way.  However, there are situations which are not well suited to this
simple model and which instead require only rechecking certain parts
of the analysis graph which has already been validated. In particular,
we consider possible untrusted \emph{updates} of a validated (trusted)
code, i.e., a code producer can (periodically) send to its consumers
new updates of a previously submitted package. We characterize the
different kind of updates, or modifications over a program.  In
particular, we include: 

\begin{enumerate}
\item the \emph{addition} of new data/predicates
and the extension of already existing predicates with new
functionalities, 
\item the \emph{deletion} of predicates or parts of them
and 
\item the \emph{replacement} of certain (parts of) predicates by new
versions for them. 
\end{enumerate}

In such a context of frequent software updates, it appears inefficient
to submit a full certificate (superseding the original one) and to
perform the checking of the entire updated program from scratch, as
needs to be done with current systems.  In the context of ACC, we
discuss the influence of the different kinds of updates on an
\emph{incremental} extension to PCC in terms of correctness and
efficiency. We also outline the main issues on the generation of
incremental certificates and the design of incremental checkers.

The paper is organized as follows. Section
\ref{sec:basics-abstr-carry} introduces briefly some notation and
preliminary notions on abstract interpretation and ACC. In Section
\ref{sec:incremental-acc}, we present a general view of incremental
ACC.  In Section \ref{cu} we describe the different kinds of updates
over a program and the way they affect the certification and checking
phases.  Section \ref{icertificates} reviews the notion of full
certificate and proposes the use of incremental certificate.  In
Section \ref{icheckers}, we discuss the extensions needed on a
non-incremental checking algorithm in order to support incrementality
and we sketch the new tasks of an incremental checking algorithm.
Finally, Section~\ref{sec:discussion} concludes.

\section{Abstraction-Carrying Code}
\label{sec:basics-abstr-carry}

Our work relies on the abstract interpretation-based analysis
algorithm of \cite{incanal-toplas} for (Constraint) Logic Programming,
(C)LP. We assume some familiarity with abstract interpretation (see
\cite{Cousot77}), (C)LP (see, e.g.,
\cite{Lloyd87,marriot-stuckey-98})  and PCC \cite{Nec97}.  

Very briefly, {\em terms} are
constructed from variables (e.g., $x$), {\em functors} (e.g., $f$) and
{\em predicates} (e.g., $p$). We denote by $\{x_1 \mapsto t_1, \ldots,
x_n \mapsto t_n\}$ the {\em substitution} $\sigma$, where $x_i \not =
x_j$, if $i \not = j$, and $t_i$ are terms. A {\em renaming} is a
substitution $\rho$ for which there exists the inverse $\rho^{-1}$
such that $\rho \rho^{-1} \equiv \rho^{-1} \rho \equiv {\it id}$. A
{\em constraint} is a conjunction of expressions built from predefined
predicates (such as inequalities over the reals) whose arguments are
constructed using predefined functions (such as real addition). An
{\em atom} has the form $p(t_1,...,t_n)$ where $p$ is a predicate
symbol and $t_i$ are terms.  A {\em literal} is either an atom or a
constraint.
A {\em rule} is of the form $H \mbox{\tt :-} D$ where $H$, the {\em
  head}, is an atom and $D$, the {\em body}, is a possibly empty
finite sequence of literals.  A {\em constraint logic program} $P \in
{\it Prog}$, or {\em program}, is a finite set of rules.  Program rules are
assumed to be normalized: only distinct variables are allowed to occur
as arguments to atoms. Furthermore, we require that each rule defining
a predicate $p$ has identical sequence of variables $x_{p_1}, \ldots
x_{p_n}$ in the head atom, i.e., $p(x_{p_1}, \ldots x_{p_n})$.  We
call this the {\em base form} of $p$. This is not restrictive since
programs can always be normalized.

An abstract interpretation-based certifier is a function {\sc
  Cer\-ti\-fi\-er}$:{\it Prog} \times \adom \times \aint \mapsto \acert$
which for a given program $P \in {\it Prog}$, an abstract domain $\dom
\in \adom$ and an abstract safety policy $\inten \in \aint$ generates
an abstract certificate $\cert \in \acert$, by using an abstract
interpreter for $\dom$, such that the certificate
entails that $P$ satisfies $\inten$. An
abstract safety policy $\inten$ is a specification of the safety
requirements given in terms of the abstract domain $\dom$. In
the following, using the same subscript
$\alpha$,  we denote that $\inten$ and $\cert$ are specifications
given as abstract semantic values of $\dom$.

The basics
for defining such certifiers (and their corresponding checkers) in ACC
are summarized in the following five points:

\begin{description}

\item[{\textbf{\em Approximation.}}] 
We consider a {\em description (or abstract)
  domain} $\langle \dom, \sqsubseteq \rangle \in \adom$ and its corresponding
\emph{concrete domain} $\langle 2^D, \subseteq \rangle$, both with a
complete lattice structure.  Description (or abstract) values and sets of concrete
values are related by an {\em abstraction} function $\alpha:
2^{D}\rightarrow \dom$, and a {\em concretization} function $\gamma:
\dom\rightarrow 2^{D}$.  
The pair $\langle \alpha, \gamma \rangle$ forms a Galois
  connection. 
 The concrete and abstract domains must be related in such a way that
  the following condition holds~\cite{Cousot77} 
  
  \[\forall x\in 2^D:~
  \gamma(\alpha(x)) \supseteq x \mbox{~~~and~~~} \forall y\in
  \dom:~ \alpha(\gamma(y)) = y \]

\noindent
In general $\sqsubseteq$ is induced by $\subseteq$ and $\alpha$.
Similarly, the operations of {\em least upper bound\/} ($\sqcup$) and
{\em greatest lower bound\/} ($\sqcap$) mimic those of $2^D$ in a
precise sense.

\medskip

\item[{\textbf{\em Analysis.}}] We consider the class of {\em fixed-point semantics}
  in which a (monotonic) semantic operator, $S_P$, is associated to
  each program $P$.
  The meaning of the program, $\lsem P \rsem$,
  is defined as the least fixed point of the $S_P$ operator, i.e.,
  $\lsem P \rsem = {\rm lfp}(S_P)$.  If $S_P$ is continuous, the least
  fixed point is the limit of an iterative process involving at most
  $\omega$ applications of $S_P$ starting from the bottom element
  of the lattice. Using abstract interpretation, we can usually only
  compute $\p_\alpha$, as $\p_\alpha={\rm lfp}(S_P^\alpha)$.
  The operator $S_P^\alpha$ is the abstract counterpart of $S_P$.

\begin{equation}\label{eq:1} {\sf analyzer}(P,\dom) = {\rm
    lfp}(S_P^\alpha) = \p_\alpha
\end{equation}
Correctness of analysis ensures that $\semabs$ safely approximates
$\sqbrack{P}$, i.e., $\sqbrack{P}\in \gamma(\sqbrack{P}_\alpha)$.
Thus, such \emph{abstraction} can be used as a certificate.

\medskip

\item[{\textbf{\em Certificate.}}]
Let $\cert$ be a safe approximation of $\semabs$.
  If an abstract safety specification $\inten$ can be proved w.r.t.\
  $\cert$, then $P$ satisfies the safety
  policy and $\cert$ is a valid certificate:

\begin{equation}\label{ecuacion2}
\cert \mbox{ is \emph{a valid certificate} for $P$ w.r.t. }
 \inten \mbox{ iff }
    \cert \sqsubseteq \inten
\end{equation}

\noindent
Note that the certificate can be stricter than the
safety specification and it is only required that $\inten$ is implied
by $\cert$.

\medskip

\item[{\textbf{\em Certifier.}}]
Together, Equations (\ref{eq:1}) and (\ref{ecuacion2}) define a certifier
which provides program fixpoints, $ \p_\alpha$, as certificates which
entail a given safety policy, i.e., by taking $\cert = \p_\alpha$.

\medskip

\item[\textbf{\em Checking.}]
  A checker is a function {\sc
    Checker}$:{\it Prog} \times \adom \times \acert \mapsto bool$
  which for a program $P \in {\it Prog}$, an abstract domain $\dom \in
    \adom$ and
  certificate $\cert \in \acert$ checks
whether $\cert$ is a fixpoint of $S^{\alpha}_P$ or not:

\begin{equation}\label{eq:3}
\small\mbox{{\sc checker}}(P,\dom,\cert) \ {\it returns\  true\ iff}\  (S^{\alpha}_P(\cert) \equiv \cert)
\end{equation}

\medskip

\item[\textbf{\em Verification Condition Regeneration.}] To retain the safety
guarantees, the consumer must regenerate a trustworthy verification
condition --Equation (\ref{ecuacion2})-- and use the incoming
certificate to test for adherence of the safety policy.

\begin{equation}\label{eq:4}
{\it  P \ is \ trusted\ iff}\  \cert \sqsubseteq \inten
\end{equation}

\noindent
A fundamental idea in ACC is that, while analysis --Equation
(\ref{eq:1})-- is an iterative process, checking --Equation~(\ref{eq:3})-- is guaranteed to be
done in a \emph{single pass} over the abstraction.

\end{description}

\section{A General View of Incremental ACC}
\label{sec:incremental-acc}

\begin{figure*}
\begin{center}
\rotatebox{-90}{
\scalebox{.53}{\includegraphics{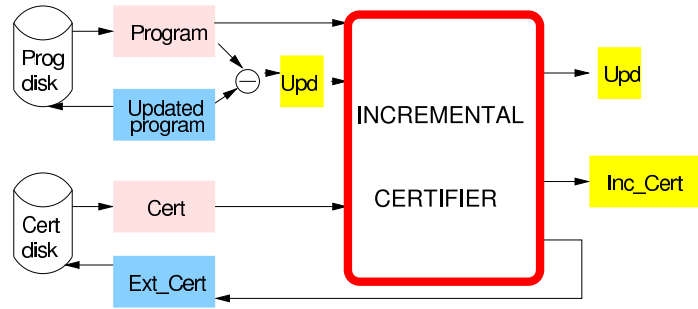}}}
\end{center}
\caption{Incremental Certification in  Abstraction-Carrying Code}
\label{inc-acc1}
\end{figure*}

\begin{figure*}
\begin{center}
\rotatebox{-90}{
\scalebox{.53}{\includegraphics{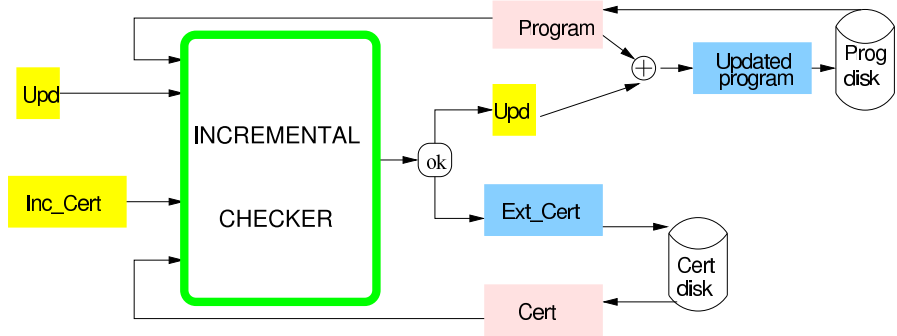}}}
\end{center}
\caption{Incremental Checking in Abstraction-Carrying Code}
\label{inc-acc2}
\end{figure*}

Figures~\ref{inc-acc1} and \ref{inc-acc2} present a general view of
the incremental certification and incremental checking processes
respectively. In Figure \ref{inc-acc1}, the producer starts from an
{\sf Updated Program}, $\eprog$, w.r.t.\ a previously certified {\sf
  Program}, $P$.  It first retrieves from disk $P$ and its
certificate, \certificateA, computed in the previous certification
phase.  Next, the process ``$\ominus$'' compares both programs and
returns the differences between them, $\Mod{P}$, i.e, the program {\sf
  Updates} which applied to $P$ results in $\eprog$, written as
$\Mod{P}=\eprog \ominus P$.  Note that, from an implementation
perspective, a program update should contain both the new updates to
be applied to the program and instructions on where to place and
remove such new code. This can be easily done by using the traditional
Unix \emph{diff} format for coding program updates. An {\sf
  Incremental Certifier} generates from \certificateA, $P$ and
$\Mod{P}$ an incremental certificate, \certificateB, which can be used
by the consumer to validate the new updates. The package
``$\Mod{P}$+\certificateB'' is submitted to the code consumer.
Finally, in order to have a compositional incremental approach, the
producer has to update the original certificate and program with the
new updates.
Thus, the resulting \certificateL\ and $\eprog$ are stored in disk
replacing \certificateA\ and $P$, respectively.

\noindent
In Figure \ref{inc-acc2}, the consumer receives the untrusted
package. In order to validate the incoming update w.r.t.\ the provided
(incremental) certificate, it first retrieves  $P$ and
\certificateA~from disk. Next, it reconstructs the updated program by using an
operator ``$\odot$'' which applies the update to $P$ and generates
$\eprog=P\odot\Mod{P}$. This can implemented by using a program in the
spirit of the traditional Unix \emph{patch} command as $\odot$
operator.  An {\sf Incremental Checker} now efficiently validates the
new modification by using the stored data and the incoming incremental
certificate. If the validation succeeds (returns {\sf ok}), the
checker will have reconstructed the full certificate.  As before, the
updated program and extended certificate are stored in disk
(superseding the previous versions) for future (incremental) updates.
In order to simplify our scheme, we assume that the safety policy and
the generation of the verification condition \cite{Nec97} are embedded
within the certifier and checker. However, in an incremental approach,
producer and consumer could perfectly agree on a new safety policy to
be implied by the modification. It should be noted that this does not
affect our incremental approach and the verification condition would
be generated exactly as in non incremental PCC.

\section{Characterization of Updates}
\label{cu}

Let us now characterize the types of updates we consider and how they
can be dealt within the ACC scheme in the context of logic
programming.  
Given a program $P$, we define an
\emph{update} of $P$, written as $\Mod{P}$, as a set of tuples of the
form $\langle A,\addition{A},\deletion{A} \rangle$, where
$A=p(x_1,\ldots,x_n)$ is an atom in base form  and:

\begin{itemize}
\item $\addition{A}$ is the set of rules which are to be added to $P$
  for predicate $p$.  This includes both the case of addition of new
 predicates, when $p$ did not exist in $P$, as well as the extension
  of additional rules (or functionality) for $p$, if it existed.

\item $\deletion{A}$ is the set of rules which are to be removed from
  $P$ for predicate $p$.
\end{itemize}

\noindent
Note that, for the sake of simplicity, we do not include the
instructions on where to place and remove such code in the
formalization of our method. We distinguish three classes of updates:
{\em addition}, {\em deletion} and {\em arbitrary changes}.

\begin{itemize}
\item The addition of predicates occurs when $\forall A,$  $\deletion{A}
= \emptyset \ \wedge \ \exists A, \ \addition{A} \not = \emptyset$. 

\item The \emph{deletion} of predicates occurs if $\forall A,$ $
\addition{A} = \emptyset\ \wedge \ \exists A, \ \deletion{A} 
\not = \emptyset$. 

\item The remaining cases are considered
\emph{arbitrary changes}.

\end{itemize}

\medskip

\begin{description}
\item[{\textbf{\em Addition of Procedures.}}]
When a program $P$ is extended with new predicates or new clauses for
existing predicates, the original certificate $\cert$ is not
guaranteed to be a fixpoint any longer, because the contribution of
the new rules can lead to a more general answer. Consider $P^{add}$
the program after applying some additions and $\certadd$ the
certificate computed from scratch for $P^{add}$. Then, $\cert
\sqsubseteq \certadd$. This means that $\cert$ is no longer valid.
Therefore, we need to perform the least upper bound (\emph{lub}) of
the contribution of the new rules and submit, together with the
extension, the new certificate $\certadd$ (or the difference of both
certificates). The consumer will thus test the safety policy w.r.t.\
$\certadd$. Consider the abstract operation ${\sf
  Alub}(\CP{}_1,\CP{}_2)$ which performs the abstract disjunction of
two descriptions. Then, we define $\certadd = {\sf
  Alub}(\cert,\sqbrack{P^{add}}_\alpha)$ and submit the incremental
certificate $\certificateA$ which is defined as the (abstract)
difference $\certadd - \cert$.  The notion of incremental certificate
is the issue of Section~\ref{icertificates}.

\medskip

\item[{\textbf{\em Deletion of Procedures.}}]
The first thing to note is that in order to entail the safety policy,
unlike extensions over the program, we need not change the certificate
at all when some predicates are deleted. Consider $P^{del}$ the
program after applying some deletions and $\certdel$ the certificate
computed from scratch for $P^{del}$.  The original certificate $\cert$
is trivially guaranteed to be a fixpoint (hence a correct
certificate), because the contribution of the rules was conjoined (by
computing the lub) to give $\cert$ and so it still correctly describes
the contribution of each remaining rule.  By applying
Equation~\ref{ecuacion2}, $\cert$ is still valid for $P^{del}$ w.r.t.
$\inten$ since $ \cert \sqsubseteq \inten$.
Therefore, more accuracy is not needed to ensure compliance with the
safety policy. This suggests that the incremental certificate can be
empty and the checking process does not have to check any predicate.
However, it can happen that a new, more precise, safety policy is
agreed by the consumer and producer. Also, this accuracy could be
required in a later modification. Although $\cert$ is a correct
certificate, it is possibly less \emph{accurate} than $\certdel$,
i.e., $\certdel \sqsubseteq \cert$.  It is therefore interesting to
define the corresponding incremental algorithm for reconstructing
$\certdel$ and checking the deletions and the propagation of their
effects.

\medskip

\item[\textbf{\em Arbitrary Changes.}]  The case of arbitrary changes
considers that rules can both be deleted from and added to an already
validated program. In this case, the new certificate for the modified
program can be either equal, more or less precise than the original
one, or even not comparable.  
Imagine that an arbitrary change replaces a rule
$R_a$, which contributes to a fixpoint $\cert^{a}$, with a new one
$R_b$ which contributes to a fixpoint $\cert^{b}$ such that
$\cert^{ab} = \alub(\cert^{a} , \cert^{b})$ and $\cert^{a} \sqsubset
\cert^{ab}$ and $\cert^{b} \sqsubset \cert^{ab}$. 
The point is that we cannot just compute an approximation of the new
rule and compute the lub with to the previous fixpoint, i.e., we
cannot use $\cert^{ab}$ as certificate and have to provide the more
accurate $\cert^{b}$.  The reason is that it might be possible to
attest the safety policy by independently using $\cert^{a}$ and
$\cert^{b}$ while it cannot be implied by using their lub
$\cert^{ab}$. This happens for certain safety policies which contain
disjunctions, i.e., $\cert^{a} \vee \cert^{b}$ does not correspond to
their lub $\cert^{ab}$.  Therefore, arbitrary changes require a
precise recomputation of the new fixpoint and its proper checking.

As a practical remark, an arbitrary update can be decomposed into an
addition and a deletion and then handled as the first cases. We have
pointed out the difference because correctness and accuracy
requirements are different in each particular case, as we have
seen above.

\end{description}

\begin{example} \label{programas}
 The next example shows a piece of a module which contains the
  following (normalized) program for the naive reversal of a list and
  uses the standard implementation of {\tt app} for appending lists:

\[\begin{array}{lll}
P_0 \equiv & \left\{
\begin{array}{lll}
({\tt rev_1}) & {\tt rev(X,Y)   :- ~X = [\ ],~ Y = [\ ].} \\
({\tt rev_2}) & {\tt rev(X,Y)   :- ~X = [U|V],~ rev(V, W),~ T = [U],~
  app(W, T, Y).} \\
({\tt app_1})&{\tt app(X,Y,Z) :-~ X = [\ ],~ Y = Z.} \\
({\tt app_4})& {\tt app(X,Y,Z) :-~ X = [U|V],~ Z = [U|W],~ app(V,Y,W)}.
\end{array}\right.
\end{array}\]

\noindent
Suppose now that the consumer modifies $P_0$ introducing two more base cases
for $\tt app$ (e.g., added automatically by a partial evaluator
\cite{pevalbook93}):

\[\begin{array}{lll}
({\tt app_2}) &{\tt app(X,Y,Z) :-~ X = [U],~ Z = [U|Y].} \\
({\tt app_3}) &{\tt app(X,Y,Z) :-~ X = [U,V],~ Z = [U,V|Y].}
\end{array}\]

\noindent
The producer must send to the consumer the set  $\Mod{P_0}$,
composed of the unique tuple:

\[\langle {\tt app(X,Y,Z)}, {\it Add}({\tt
  app(X,Y,Z)}), {\it Del}({\tt app(X,Y,Z)}) \rangle\]

\noindent
where $\addition{{\tt app(X,Y,Z)}} = \{{\tt app_2},{\tt app_3}\}$ and
$\deletion{{\tt app(X,Y,Z)}} = \emptyset$, i.e., we are in the case of
an {\em addition} of predicates only. Let us name $P_1$ to the program
resulting from adding rules ${\tt app_2}$ and ${\tt app_3}$ to $P_0$.
Note that these rules do not add any further information to the
program (i.e., the certificate for $P_0$ and $P_1$ would remain the
same and, as we will see, the {\em incremental certificate is empty}).

Consider now the following new definition
for predicate {\tt app} which is a specialization of the previous
  {\tt app} to concatenate lists of {\tt a}'s of the same length:

\[\begin{array}{lll}
{\tt (Napp_1)} &{\tt app(X,Y,Z) :-~ X = [\ ],Y = [\ ], Z = [\ ].}\\
{\tt (Napp_2)} & {\tt app(X,Y,Z) :-~ X = [a|V],Y=[a|U], Z = [a,a|W],
     app(V,U,W)}.
\end{array}\]

\noindent
The update consists in deleting all rules for predicate {\tt app} in
$P_1$ and replacing them by ${\tt Napp_1}$ and
${\tt Napp_2}$. Let $P_2$ be the resulting program.
$\Mod{P_1}$ is composed again of a unique tuple with
the following information:

\[ \begin{array}{lll}
\addition{{\tt app(X,Y,Z)}}=\{{\tt Napp_1}, {\tt Napp_2}\}\\
\deletion{{\tt app(X,Y,Z)}}= {\tt \{app_1,app_2,app_3,app_4\}}
\end{array} \]

\noindent
In this case, we are in presence of an {\em arbitrary change}, and as
we will show in Example \ref{ultimo}, the {\em incremental
  certificate} will not be empty in this case (since by using the
abstract domain  $\defff$ in Example \ref{ex:dom-call}, the
fixpoint for $P_2$ will change w.r.t. the one for $P_1$).  \hfill
$\Box$
\end{example}

\section{Incremental Certificates}
\label{icertificates}

Although ACC and incremental ACC, as outlined above, are general
proposals not tied to any particular programming paradigm, our
developments for incremental ACC (as well as for the original ACC
framework \cite{lpar04-ai-safety}) are formalized in the context of
(C)LP. A main idea in ACC \cite{lpar04-ai-safety} is that a {\em
  certificate}, $\certificateA$, is automatically generated by using
the \emph{complete} set of {\em entries} returned by an abstract
fixpoint analysis algorithm. For concreteness, we rely on an abstract
interpretation-based analysis algorithm in the style of the generic
analyzer of \cite{incanal-toplas}. 

The analysis algorithm of \cite{incanal-toplas}, which we refer to as
\analyzerA, given a program $P$ and an abstract domain
$\dom$, receives a set of {\em
  call patterns} $\atom \in \aatom$ (or Abstract Atoms)
   which are a description of
the calling modes into the program, and constructs an \emph{analysis
  graph}~\cite{bruy91} for $\atom$ which is an \emph{abstraction} of
the (possibly infinite) set of (possibly infinite) trees explored by
the concrete execution of initial calls described by $\atom$ in $P$.
Formally, a \emph{call pattern} $A:{\it CP} \in \aatom$ is composed of an atom in
base form, $A \equiv p(X_1,\ldots, X_n)$, and a description
in the abstract domain, ${\it CP}$, for $A$.

The program analysis graph computed by \analyzerA$(\atom)$ for $P$ in
$\dom$ can be implicitly represented by means of two data structures,
the {\em answer table} (${\it AT}$) and the {\em dependency arc table}
(${\it DAT}$), which are the
output of the algorithm  \analyzerA.
Each element (or {\em entry})
in the answer table takes the form $A:{\it CP} \mapsto {\it AP}$ such
that, for the atom $A$, ${\it CP}$ is its \emph{call} description and
${\it AP}$ its \emph{success} (or answer) description in the abstract
domain. Informally, such entry should be interpreted as ``the answer
pattern for calls to $A$ satisfying precondition ${\it CP}$ accomplishes
postcondition $AP$''.  The
dependency arc table is not relevant now, although it is
fundamental in the design of the incremental checking, as we will see
later.  All the details and the formalization of the algorithm can be
found in \cite{incanal-toplas}.

Our proposal for the incremental checking is that,
if the consumer keeps the original (fixed-point)
abstraction $\certificateA$, it is possible to provide only the
program updates and
the incremental certificate $\certificateB$. Concretely, given:

\begin{itemize}
\item an update $\Mod{P}$ of $P$,
\item the certificate \certificateA~for
$P$ and $\atom$,
\item the certificate \certificateL~for
$P\odot\Mod{P}$ and $\atom$
\end{itemize}

\noindent
we define \certificateB, the
\emph{incremental certificate} for $\Mod{P}$ w.r.t.\ $\certificateA$,
as the difference of certificates $\certificateL$ and $\certificateA$,
i.e., the set of entries in $\certificateL$ not occurring in $\certificateA$.
The first obvious advantage is that the size of the
transmitted certificate can be considerably reduced.  Let us see an example.

\begin{example}\label{ex:dom-call} Consider program $P_0$ in Example
  \ref{programas}. The description domain that we are going to use in our examples
  is the {\em
    definite Boolean functions}~\cite{pos-def94}, denoted $\defff$.
  The key idea in this description is to use implication to capture
  groundness dependencies.  The reading of the function $x \rightarrow
  y$ is ``if the program variable $x$ is (becomes) ground, so is
  (does) program variable $y$.''  For example, the best description of
  the constraint ${\tt f(X,Y) = f(a,g(U,V))}$ is ${\tt X \wedge (Y
  \leftrightarrow (U \wedge V))}$.
  The most general description $\tt \top$ does not provide information
  about any variable.  The least general substitution $\tt \bot$
  assigns the empty set of values to each variable.
For the analysis of our running example, we consider the set of
 call
  patterns $\atom=\{{\tt rev(X,}$ ${\tt Y):\top}\}$, i.e., no entry information
  is provided on $\tt X$ nor $\tt Y$. \analyzerA(\{${\tt rev(X,
    Y):\top}$\})~returns in the answer table, ${\it AT}$, the following entries:

\[ \begin{array}{ll@{~~~~~~}llllll}
({\it A_1})& \ \ \entry{\tt rev(X, Y) : \top }{ \tt X \leftrightarrow
  Y}  \\
({\it A_2})& \ \ \entry{\tt app(X, Y, Z) : \top}{\tt  (X \wedge Y)
  \leftrightarrow Z}
\end{array} \]

\noindent
The certificate
\certificateA~for this example is then composed of the entries ${\it
  A_1}$ and ${\it A_2}$.
Consider now the addition of rules $\tt app_2$ and $\tt app_3$ in
$P_0$, i.e., program $P_1$ of Example \ref{programas}.
The analysis algorithm of \cite{incanal-toplas}
returns as $\certificateL$ the same answer table ${\it AT}$ as for
$P_0$, since the added
rules do
not affect the fixpoint result, i.e., they do not add any further
information. Thus, the incremental certificate
\certificateB~associated to such an update is empty.  \hfill $\Box$

\end{example}

\section{Incremental Checking}
\label{icheckers}
Intuitively, an  abstract interpretation-based checking algorithm
(like the one in \cite{lpar04-ai-safety})
receives as input a program $P$, a set of abstract atoms $\atom$ and a
certificate \certificateA\, and constructs a program analysis graph in
a single iteration by assuming the fixpoint information in
\certificateA.
While the graph is being constructed,
the obtained answers are stored in an answer table ${\it AT_{\it
    mem}}$ (initially empty)
and compared with the corresponding fixpoints stored in \certificateA.
If any of the computed answers is not consistent with the certificate
(i.e., it is greater than the fixpoint), the certificate is considered
invalid and the program is rejected.
Otherwise, \certificateA~gets accepted 
and ${\it AT_{\it mem}} \equiv \certificateA$.

 \subsection{Checking with Dependencies}

 In order to define an incremental checking, the checking algorithm in
 \cite{lpar04-ai-safety} needs to be modified to compute (and store)
 also the dependencies between the atoms in the answer table.  In
 \cite{inc-acc-tr}, we have instrumented a checking algorithm for full
 certificates with a \emph{Dependency Arc Table}.  This structure, ${\it DAT}$, is not required by non
 incremental checkers but it is fundamental to support an incremental
 design.
The ${\it DAT}$ returned by \analyzerA~is composed of
 arcs (or {\em dependencies}) of the form $A_k:{\it CP} \Rightarrow
 B_{k,i}:{\it CP}'$ associated to a program rule $A_k \mbox{ :- }
 B_{k,1},\ldots,B_{k,n}$ with $i \in \{1,..n\}$, where $ B_{k,i}$ is an atom.  The intended meaning
 of such a dependency is that the answer for $A_k:{\it CP}$ depends on the
 answer for $B_{k,i}:{\it CP}'$, say ${\it AP}$.  Thus, if ${\it AP}$ changes with
 the update of some rule for $B_{k,i}$ then, the arc $A_k:{\it CP}
 \Rightarrow \ B_{k,i}:{\it CP}'$ must be reprocessed in order to compute
 the new answer for $A_k:{\it CP}$. This is to say that the rule for $A_k$
 has to be processed again starting from atom $B_{k,i}$, i.e., we do
 not need to process the part  $A_k \mbox{ :- }
 B_{k,1},\ldots,B_{k,i-1}$ because it is not affected by the changes.

 In the following, we assume that {\sc checker} is a non incremental
 checker such that, if the call {\sc checker}($P,\atom,\certificateA$)
 does not fail, then it returns the reconstructed answer table ${\it
   AT_{\it mem}}$ and the set of dependencies ${\it DAT_{\it mem}}$
 which have been generated. In such a case, we say that $\certificateA$
 has been {\em checked} or {\em accepted}. By the correctness of the checker
 \cite{lpar04-ai-safety}, the reconstructed structures contain exactly
 the same data as the answer table and the dependency arc table
 computed by the analysis algorithm \analyzerA($\atom$) for the
 program $P$.

\begin{example}\label{ex:dom-call1} Consider the program $P_0$ in
  Example \ref{programas}. \analyzerA~returns, together with ${\it
    AT}$, the following dependency arc table:

\[ \begin{array}{llllllll}
 ({\it D_1}) & \dependency{\tt rev(X,Y):\top}{\tt rev(V,W):\top} \\
 ({\it D_2}) & \dependency{\tt
  rev(X,Y):\top}{\tt app(W,T,Y):\top} \\
({\it D_3}) & \dependency{\tt app(X,Y,Z):\top}{\tt app(V,Y,W):\top}
\end{array} \]

\noindent
Intuitively, $D_2$ denotes that the answer for ${\tt rev(X,Y):\top}$
may change if the answer for ${\tt app(W,T,Y):\top}$ changes. In such
a case, the second rule $\tt rev_2$ for ${\tt rev}$ must be processed
again starting from atom ${\tt app(W,T,Y)}$ in order to recompute the
fixpoint for ${\tt rev(X,Y):\top}$.  $D_1$ and $D_3$ reflect the
recursivity of ${\tt rev(X,Y):\top}$ and ${\tt app(W,}$ ${\tt
  T,Y):\top}$, respectively, since they depend on themselves (rules
$\tt rev_2$ and $\tt app_4$ respectively).  The detailed steps
performed by the algorithm can be found in \cite{incanal-toplas}. Note
that, the {\sc checker} executed for the call pattern at hand,
computes (and stores) in ${\it AT_{\it mem}}$ the entries $A_1$, $A_2$
in Example \ref{ex:dom-call}, and, after traversing rules $\tt rev_2$
and $\tt app_4$, it stores in ${\it DAT_{\it mem}}$ the dependencies
$D_1$, $D_2$ and $D_3$.  \hfill $\Box$

\end{example}

\subsection{Additional Tasks of an Incremental Checker}
\noindent
In order to support incrementality, the final values of the data
structures ${\it AT_{\it mem}}$, ${\it DAT_{\it mem}}$ and $P$ must be
available after the end of the execution of the checker. Thus, we
denote by ${\it AT_{\it persist}}$, ${\it DAT_{\it persist}}$ and
${\it P}_{persist}$ the copy in persistent memory (i.e., in disk) of
such structures. Now, we outline in a very general way the additional
tasks that  an incremental checking algorithm
({\sc inc\_check} in the following) has to perform. The complete code
of the algorithm can be found in \cite{inc-acc-tr}.
It receives as input
parameters an update $\Mod{P}$ of the original program $P$, a set of
abstract atoms  $\atom \in \aatom$ and the incremental certificate
$\certificateB$ for $\Mod{P}$ w.r.t. $\certificateA$. The following
tasks are carried out by an incremental checker:

\begin{description}

\item[{\bf Step 1)}] It retrieves from memory ${\it AT_{mem}}:={\it AT_{persist}}$,
  ${\it DAT}_{\it mem}$ $:={\it DAT}_{\it persist}$ and $P:=
  P_{persist}$ (stored in persistent memory in a previous checking
  phase) and generates $P_{\it mem}:=P \odot \Mod{P}$.

\medskip

\item[{\bf Step 2)}] It rechecks all entries in
${\it AT_{mem}}$ which have been directly affected by an
update. Concretely, for each $A:{\it CP} \in {\it AT_{mem}}$, such
that $A$ has an entry in $\Mod{P}$, a call to {\sc checker}$(P \odot
\Mod{P}, \{A:{\it CP}\}, \certificateB)$ is generated, marking the
entry as {\em checked} (or {\em accepted}) $A:{\it
  CP}^{\it check}$. This guarantees that the incremental checking
process is done in one pass (i.e., rules used to check $A:{\it CP}$
are traversed at most once).

\medskip

\item [{\bf Step 3)}] It propagates and rechecks the indirect effect of these
  changes by inspecting the dependencies in ${\it DAT_{\it mem}}$.
  Thus, for all $A:{\it CP}^{check} \in \certificateB$, if there
  exists a dependency of the form $B:{\it CP_B} \Rightarrow A:{\it
    CP}$ (modulo renaming) in ${\it DAT_{\it mem}}$ such that $B:{\it
    CP_B}$ is not marked as checked, then a call to {\sc checker}$(P
  \odot \Mod{P}, \{B:{\it CP_B}\}, \certificateB)$ is generated and
  $B:{\it CP}_B$ is marked as checked. This process is repeated until
  there are no dependencies satisfying the above condition. Note
  that the condition $A:{\it CP}^{check} \in \certificateB$ ensures
  that the answer for $A:{\it CP}$ has changed w.r.t. $\certificateA$.
  Otherwise nothing has to be done (this will allow us to reduce the
  checking time w.r.t a full checking process for $P$ and
  $\certificateL$).

\medskip

\item[{\bf Step 4)}]  If it does not issue an \error~then it removes
from ${\it AT}_{\it mem}$ those entries
corresponding to deleted rules. We can identity them by exploring
${\it DAT_{\it mem}}$. Concretely, for all $A:{\it CP} \in {\it
  AT}_{\it mem}$, $A:{\it CP} \not \in S_{\it \alpha}$, if
there not exists a dependency $B:{\it CP'} \Rightarrow A:{\it CP}$ in
${\it DAT_{\it mem}}$ then remove $A:{\it CP}$ from ${\it AT_{\it
    mem}}$.

\medskip

\item[{\bf Step 5)}] It stores
 ${\it AT_{\it persist}}:={\it AT_{\it
      mem}}$,  ${\it DAT_{\it persist}}$ $:={\it DAT_{\it mem}}$ and
${\it P_{\it persist}}:=P_{\it mem}$.

\end{description}
Our first example is intended to illustrate a situation in which the
task performed by the incremental checker can be optimized such that it
only checks a part of the analysis graph.

\begin{example} \label{exampleCh1}
Consider the addition of rules $\tt app_2$ and $\tt app_3$ to program
$P_0$, which results in program $P_1$ (Example
\ref{programas}). As shown in Example \ref{ex:dom-call},
the incremental certificate
\certificateB~associated to such an update is empty.
The incremental checking
algorithm {\sc inc\_check} proceeds as follows:

\begin{description}
\item[{\bf Step 1)}]  ${\it AT}_{\it mem}$ and
${\it DAT}_{\it mem}$ are initialized with $A_1$,$A_2$ (Example
\ref{ex:dom-call})  and $D_1$,$D_2$
and $D_3$ (Example \ref{ex:dom-call1}) respectively.
$P_{\it mem} \equiv P_1$.

\medskip

\item[{\bf Step 2)}] Since ${\tt
  app(X,Y,Z):\top}
\in {\it AT_{\it mem}}$ and $\addition{{\tt app(X,Y,Z)}}$ is not empty, then
a call to {\sc checker}($P_1$,\{$\tt app(X,Y,Z):\top$\},$\certificateB$)
is generated
 in order to ensure that the fixpoint is preserved. Now, $\tt
 app(X,Y,Z):\top$ is marked as checked.

\medskip

\item[{\bf Step 3)}] No checking has to be done since $\certificateB$
  is empty.

\medskip

\item[{\bf Step 4)}] Nothing is done since $\tt app(X,Y,Z):\top$
  occurs at the right-hand side of dependency $D_3$.

\medskip

\item[{\bf Step 5)}] Finally,
once \certificateB~has been validated, the
consumer stores the answer table ${\it AT_{\it mem}}$, the dependency
arc table ${\it DAT_{\it mem}}$ and
the program $P_{\it mem}$ in disk with the same values as in
Step 1. \hfill $\Box$
\end{description}
\end{example}

\noindent
Our second example is intended to show how to propagate the effect of
a change to the part of the analysis graph indirectly affected by such update.

\begin{example} \label{ultimo}
The update consists in deleting all rules for predicate {\tt app} in
program $P_1$ of Example~\ref{programas} (which results in program
$P_2$),  and replacing them
by ${\tt Napp_1}$ and
${\tt Napp_2}$. After running the \analyzerA~for $P_2$,
the following answer table and dependencies are computed:

\[ \begin{array}{llllllll}
{\it (NA_1)}\ \ \entry{\tt rev(X, Y) : \top}{\tt X \wedge Y}  \\
{\it (NA_2)}\ \ \entry{\tt app(X, Y,Z) : \top}{\tt X \wedge Y \wedge Z}\\
{\it (NA_3)}\ \ \entry{\tt app(X, Y,Z) : X}{\tt X \wedge Y \wedge Z} \\
{\it (ND_1)}\ \ \dependency{\tt rev(X,Y):\top}{\tt rev(V,W):\top} \\
{\it (ND_2)}\ \   \dependency{\tt rev(X,Y):\top}{\tt app(W,T,Y):W} \\
{\it (ND_3)}\ \  \dependency{\tt app(X,Y,Z):X}{\tt app(V,U,W):V}
\end{array} \]

\noindent
Note that the
analysis information has changed because the new
definition of {\tt app} allows inferring that all its arguments are
ground upon success ${\it (NA_2}$ and ${\it NA_3)}$.\footnote{Note hat
${\it NA_3}$ is subsumed by ${\it NA_2}$ and we could indeed only
submit  ${\it NA_2}$. The incremental checking algorithm should be
modified to search entries which are equal or more general than the
required one.} This change
propagates to the answer of {\tt rev} and allows inferring that,
regardless of the call pattern, both arguments of {\tt rev} will be
ground on the exit ${\it (NA_1)}$. The incremental
certificate \certificateB~contains ${\it NA_3}$ as it corresponds to a
new call pattern and contains also ${\it NA_1}$ and
${\it NA_2}$ since their answers have changed w.r.t. the ones stored in
$\certificateA$ (Example~\ref{ex:dom-call}).
Let us illustrate the incremental checking process carried out to
validate this update.

\begin{description}
\item[{\bf Step 1)}] We retrieve
from disk the answer table, dependency arc table and the program
stored in Step 5 of Example \ref{exampleCh1}. Now $P_{\it mem} \equiv P_2$.

\medskip

\item[{\bf Step 2)}] Similar to Step 2 of Example  \ref{exampleCh1},
  but considering the new rules for $\tt app$.

\medskip

\item[{\bf Step 3)}] Since we have the dependency $D_2 \in {\it DAT_{\it mem}}$ and
${\tt app(X,Y,Z):\top} \in \certificateB$,
a call to {\sc checker}($P_2$,$\{\tt rev(X,Y):\top\}$,$\certificateB$) is
generated to ensure that the new fixpoint for $\tt rev(X,Y):\top$ is
valid. In the checking process, when traversing the rule $\tt rev_2$,
the new call pattern $\tt app(X,Y,Z):X$ occurs and it is also
validated by calling to {\sc checker}. When traversing rule $\tt
Napp_2$, the dependency $D_3$ is replaced by the new one ${\it ND_3}$
in ${\it DAT_{\it mem}}$, and the call pattern is marked as checked.
Similarly, the
dependency ${\it D_2}$ is replaced by
the new one ${\it ND_2}$ and
 $\tt rev(X,Y):\top$
is marked as checked. Now, all call patterns have been checked and the
process finishes.

\medskip

\item[{\bf Step 4)}] The entry ${\it NA}_2$ is removed from ${\it
    AT_{\it mem}}$ since it does not occur at the right-hand side of
  any dependency.

\medskip

\item[{\bf Step 5)}] The
consumer stores the answer table ${\it AT_{\it mem}}:=\{{\it
  NA_1,NA_3}\}$,
the dependency
arc table ${\it DAT_{\it mem}}:=\{{\it ND_1,ND_2,ND_3}\}$ and
the program $P_{\it mem} := P_2$ in disk. \hfill $\Box$

\end{description}
\end{example}
The definition of the algorithm {\sc inc\_check} can be found in
\cite{inc-acc-tr}, together with the proof of the correctness of the
algorithm. Informally, correctness amounts to saying that if {\sc
  inc\_check} does not issue an error, then it returns as computed
answer table the extended certificate $\certificateL$ for the updated
program.  Moreover, we ensure that it does not iterate during the
reconstruction of any answer.

\section{Conclusions}\label{sec:discussion}

Our proposal to incremental ACC aims at reducing the size of
certificates and the checking time when a supplier provides an
untrusted update of a (previously) validated package.  Essentially,
when a program is subject to an update, the incremental certificate we
propose contains only the \emph{difference} between the original
certificate for the initial program and the new certificate for the
updated one.  Checking time is reduced by traversing only those parts
of the abstraction which are affected by the changes rather than
the whole abstraction. 
An important point to note is that our incremental
approach requires the original certificate and the dependency arc
table to be stored by the consumer side for upcoming updates.  The
appropriateness of using the incremental approach will therefore
depend on the particular features of the consumer system and the
frequency of software updates.  In general, our approach seems to be
more suitable when the consumer prefers to minimize as much as
possible the waiting time for receiving and validating the certificate
while storage requirements are not scarce.  We believe that, in
everyday practice, time-consuming safety tests would be avoided by
many users, while they would probably accept to store the safety
certificate and dependencies associated to the package. Nevertheless,
there can sometimes be situations where storage resources can be very
limited,
while runtime resources for performing upcoming checkings could still
be sufficient.
We are now in the process of extending the ACC implementation already
available in the \ciaopp\ system to support incrementality. Our
preliminary results in certificate reduction are very promising. 
We
expect optimizations in the checking time similar to those achieved in
the case of incremental analysis (see, e.g., \cite{incanal-toplas}).

\section*{Acknowledgments}

This work was funded in part by the Information Society Technologies
program of the European Commission, Future and Emerging Technologies
under the IST-15905 {\em MOBIUS} project, by the Spanish Ministry of
Education under the TIN-2005-09207 {\em MERIT} project, and the Madrid
Regional Government under the S-0505/TIC/0407 \emph{PROMESAS} project.
The authors would like to thank the anonymous referees of WLPE for
their useful comments.

\bibliographystyle{plain}

\begin{thebibliography}{10}

\bibitem{inc-acc-tr}
E.~Albert, P.~Arenas, and G.~Puebla.
\newblock An {I}ncremental {A}pproach to {A}bstraction-{C}arrying {C}ode.
\newblock Technical Report CLIP3/2006, Technical University of Madrid (UPM),
  School of Computer Science, UPM, March 2006.

\bibitem{lpar04-ai-safety}
E.~Albert, G.~Puebla, and M.~Hermenegildo.
\newblock {A}bstraction-{Carrying} {Code}.
\newblock In {\em Proc. of LPAR'04}, number 3452 in LNAI, pages 380--397.
  Springer-Verlag, 2005.

\bibitem{pos-def94}
T.~Armstrong, K.~Marriott, P.~Schachte, and H.~S{\o}ndergaard.
\newblock Boolean functions for dependency analysis: Algebraic properties and
  efficient representation.
\newblock In Springer-Verlag, editor, {\em Static Analysis Symposium, SAS'94},
  number 864 in LNCS, pages 266--280, Namur, Belgium, September 1994.

\bibitem{bruy91}
M.~Bruynooghe.
\newblock {A} {P}ractical {F}ramework for the {A}bstract {I}nterpretation of
  {L}ogic {P}rograms.
\newblock {\em Journal of Logic Programming}, 10:91--124, 1991.

\bibitem{Cousot77}
P.~Cousot and R.~Cousot.
\newblock {A}bstract {I}nterpretation: a {U}nified {L}attice {M}odel for
  {S}tatic {A}nalysis of {P}rograms by {C}onstruction or {A}pproximation of
  {F}ixpoints.
\newblock In {\em {F}ourth {ACM} {S}ymposium on {P}rinciples of {P}rogramming
  {L}anguages}, pages 238--252, 1977.

\bibitem{incanal-toplas}
M.~Hermenegildo, G.~Puebla, K.~Marriott, and P.~Stuckey.
\newblock {I}ncremental {A}nalysis of {C}onstraint {L}ogic {P}rograms.
\newblock {\em ACM Transactions on Programming Languages and Systems},
  22(2):187--223, March 2000.

\bibitem{pevalbook93}
N.D. Jones, C.K. Gomard, and P.~Sestoft.
\newblock {\em {P}artial {E}valuation and {A}utomatic {P}rogram {G}eneration}.
\newblock Prentice Hall, New York, 1993.

\bibitem{JVM03}
Xavier Leroy.
\newblock Java bytecode verification: algorithms and formalizations.
\newblock {\em Journal of Automated Reasoning}, 30(3-4):235--269, 2003.

\bibitem{Lloyd87}
J.W. Lloyd.
\newblock {\em Foundations of Logic Programming}.
\newblock Springer, second, extended edition, 1987.

\bibitem{marriot-stuckey-98}
Kim Marriot and Peter Stuckey.
\newblock {\em {P}rogramming with {C}onstraints: {A}n {I}ntroduction}.
\newblock The MIT Press, 1998.

\bibitem{Nec97}
G.~{}Necula.
\newblock Proof-{C}arrying {C}ode.
\newblock In {\em Proc. of POPL'97}, pages 106--119. ACM Press, 1997.

\bibitem{RR98}
K.~Rose, E.~Rose.
\newblock Lightweight bytecode verification.
\newblock In {\em OOPSLA Workshop on Formal Underpinnings of Java}, 1998.

\end{thebibliography}

\end{document}